\documentclass[floats,floatfix,showpacs,prd,twocolumn,superscriptaddress,nofootinbib,nolongbibliography,reprint]{revtex4-2}
\def\apj{\rm{ApJ}}
\def\apjl{\rm{ApJ}}

\def\mnras{\rm{MNRAS}}
\def\prd{\rm{PRD}}

\def\prx{\rm{PRX}}
\def\pasa{\rm{PASA}}

\usepackage{amssymb}
\usepackage{amsmath}
\usepackage{bbm}
\usepackage{graphicx} 
\usepackage[dvipsnames, usenames]{xcolor}
\definecolor{linkcolor}{rgb}{0.0,0.3,0.5}
\usepackage[unicode, colorlinks=true, linkcolor=linkcolor, citecolor=linkcolor, filecolor=linkcolor,urlcolor=linkcolor, pdfusetitle]{hyperref}
\usepackage{microtype} 

\newcommand{\PE}{\ensuremath{\textsc{pe}}}
\newcommand{\POP}{\ensuremath{\textrm{pop}}}
\newcommand{\DET}{\ensuremath{\mathrm{det}}}
\newcommand{\Nobs}{\ensuremath{N_\mathrm{obs}}}
\newcommand{\bigN}{\ensuremath{[\![1,\Nobs]\!]}}
\newcommand{\alltheta}{\ensuremath{\{\theta_i\}}}
\newcommand{\alldata}{\ensuremath{\{d_i\}}}
\newcommand{\mosttheta}{\ensuremath{\{\theta_{i\neq j}\}}}
\newcommand{\mostdata}{\ensuremath{\{d_{i\neq j}\}}}
\newcommand{\dint}{\ensuremath{\mathrm{d}}}
\renewcommand{\sim}{\mathchar"5218\relax\,}

\newcommand{\bham}{
    \affiliation{Institute for Gravitational Wave
    Astronomy \& School of Physics and Astronomy, 
    University of Birmingham, Edgbaston, Birmingham 
    B15 2TT, UK}
}

\newcommand{\milan}{
    \affiliation{Dipartimento di Fisica ``G. Occhialini'', Universit\'a degli Studi di Milano-Bicocca, Piazza della Scienza 3, 20126 Milano, Italy}
    \affiliation{INFN, Sezione di Milano-Bicocca, Piazza della Scienza 3, 20126 Milano, Italy}
}

\begin{document}

\title{Population-informed priors in gravitational-wave astronomy}

\author{Christopher J.\ Moore} \email{moorecj@bham.ac.uk} \bham
\author{Davide Gerosa} \bham \milan

\date{\today}

\begin{abstract}
We describe a Bayesian formalism for analyzing individual gravitational-wave events in light of the rest of an observed population. 
This analysis reveals how the idea of a ``population-informed prior'' arises naturally from a suitable marginalization of an underlying hierarchical Bayesian model which consistently accounts for selection effects. 
Our formalism naturally leads to the presence of ``leave-one-out'' distributions which include subsets of events. 
This differs from other approximations, also known as empirical Bayes methods, which effectively double count one or more events. 
We design a double-reweighting post-processing strategy that uses only existing data products to reconstruct the resulting population-informed posterior distributions. 
Although the correction we highlight is an important conceptual point, we find it has a limited impact on the current catalog of gravitational-wave events. 
Our approach further allows us to study, for the first time in the gravitational-wave literature, correlations between the parameters of individual events and those of the population.  
\end{abstract}

\maketitle


\begin{figure*}
\begin{minipage}[t]{0.49\linewidth}
\includegraphics[page=1,width=0.35\textwidth]{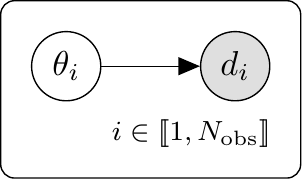} 
\vspace{0.3cm}\\
{\normalsize {\bf (a)} Single-event parameter estimation} \\[10pt]
\end{minipage}
\hfill
\begin{minipage}[t]{0.49\linewidth}
\includegraphics[page=2,width=0.50\textwidth]{pgm_only}
\vspace{0.3cm}\\
{\normalsize {\bf(b)} Population inference} \\[10pt]
\end{minipage}

\vspace{0.6cm}

\begin{minipage}[t]{0.49\linewidth}
\includegraphics[page=3,width=0.64\textwidth]{pgm_only}
\vspace{0.3cm}\\
{\normalsize {\bf (c)}  Population-informed single-event inference} \\[10pt]
\end{minipage}
\hfill
\begin{minipage}[t]{0.49\linewidth}
\includegraphics[page=4,width=0.5\textwidth]{pgm_only}
\vspace{0.3cm}\\
{\normalsize {\bf(d)} Full hierarchical model} \\[10pt]
\end{minipage}
\vspace{0.2cm}
    \caption{ \label{pgms}
    Probabilistic graphical models (PGMs) for the analyses described in the introduction. Observed data are indicated with gray circles; empty circles indicate model parameters one wishes to sample; hatched circles indicate model parameters that are marginalized over in the analysis; boxes indicate parts of the analysis that are repeated independently for multiple events; diamonds indicate steps in the analysis where selection effects must be taken into account. 
    \emph{Panel (a): Single-event parameter estimation.} 
    It is common practice for every event to be analyzed individually under an uninformative choice of prior, $\pi_\PE(\theta)$; the target distribution for this analysis is given in Eq.~(\ref{eq:event_posterior}).
    \emph{Panel (b): Population inference.} 
    When interested solely in the population parameters, these can be inferred using posterior samples from the single-event analyses; the target distribution for this analysis is given in Eq.~(\ref{eq:population_posterior_marg}). The $\theta_i$ parameters are analytically marginalized over and therefore cannot be inferred from this analysis. Note how the PGM from panel (a) is contained within this diagram.
    \emph{Panel (c): Population-informed single-event inference.} 
    In this case, single events are analyzed in light of the whole population; the target distribution for this analysis is given in Eq.~(\ref{eq:pop_informed_prior}). Note how the PGM from panel (b) is contained within this diagram but with one event omitted and $\lambda$ marginalised out.
    \emph{Panel (d): Full hierarchical model.} 
    In principle, the population can be analyzed simultaneously with all of the events in a full hierarchical Bayesian model; the target distribution is given in Eq.~(\ref{eq:hierarchical}). The analyses in panels (b) and (c) can be obtained by marginalising this over $\alltheta$ (see Sec.~\ref{subsec:population_inference}) and $(\mosttheta,\lambda)$ (see Sec.~\ref{subsec:pop_informed_priors}), respectively.
    } 
\end{figure*}


%
%
\section{Introduction} \label{sec:intro}

Bayesian statistics plays a prominent role in gravitational-wave (GW) astronomy, where it is routinely used to infer the properties of individual binary black hole (BH) events \cite{2019PhRvX...9c1040A,2021PhRvX..11b1053A}.
Bayesian statistics is also used to infer the properties of the underlying distribution of sources~\cite{2019ApJ...882L..24A,2021ApJ...913L...7A}, assuming all events come from the same modeled population (which can be a mixture of several channels; for reviews see \cite{2019PASA...36...10T,2020arXiv200705579V}). 
Although individual-event and population inferences are often treated separately for practical and computational purposes, they can be viewed as two sides of the same coin: namely a full, hierarchical Bayesian model. 

Hierarchical Bayesian models have been successfully applied to many astronomical data sets, including spectroscopic data for the determination of stellar ages~\cite{2016ApJ...817...40F}, light curve~\cite{2014ApJ...796...47M} and radial velocity~\cite{2010ApJ...725.2166H} data for the determination of exoplanet obliquities and eccentricities respectively, and astroseismic data for the determination of stellar inclinations~\cite{2016ApJ...819...85C} and helium enrichment~\cite{2021MNRAS.tmp.1343L}.
One benefit of hierarchical Bayesian models is that one can obtain improved measurements of the parameters of individual events by exploiting the fact that they are part of a large catalog---an approach that can be described as using a \emph{population-informed prior}. 
Taking the first GW event as an example, this line of reasoning is equivalent to asking: %
\begin{quote}
What can be learned about GW150914 using not only the $\sim 0.2\,$s of data from September 14th, 2015, but rather from all of the LIGO/Virgo observations to date? 
\end{quote}

We now describe several different, but related analyses.
The interplay between these procedures can be visualized using the probabilistic graphical models (PGMs) in Fig.~\ref{pgms}.
In these diagrams, data and parameters are indicated with circles while arrows represent conditional probabilistic dependencies between quantities.
Figure~\ref{pgms}(a) illustrates the standard single-event parameter estimation (e.g.~\cite{2019PhRvX...9c1040A,2021PhRvX..11b1053A}): one selects an event (index $i$) from the catalog and infers its parameters $\theta_i$ (masses, spins, etc.) using only the data $d_i$ for that event and an uninformative prior. 
Figure~\ref{pgms}(b) illustrates the standard population analysis (e.g.~\cite{2019ApJ...882L..24A,2021ApJ...913L...7A}): hyperparameters $\lambda$ describing the source population (e.g. the slope of the mass spectrum) are inferred using the results of all of the single-event analyses, taking care to properly account for selection effects. In this approach, the single-event parameters are marginalized over and, therefore, cannot be sampled.
Figure~\ref{pgms}(c) illustrates a population-informed single-event analysis, which is the main topic of this paper. 
Targeting event $j$, this analysis uses the data for all other events, $\mostdata$, to infer the population parameters $\lambda$ which are then marginalized over, while incorporating the data $d_j$, to infer the parameters $\theta_j$.
As is shown below, the analyses depicted in panels (b) and (c) are appropriate marginalizations of the full hierarchical Bayesian model indicated in Fig.~\ref{pgms}(d).

This paper describes a complete formalism for performing population-informed single-event analyses in the presence of selection effects, showing in particular how it follows from a full hierarchical Bayesian model (Sec.~\ref{sec:inference}). 
Our solution differs from some previous studies that rely on heuristic derivations and
that sometimes mistreat selection effects and/or implicitly apply an empirical Bayes method \cite{casella1985introduction}---a known approximation of a hierarchical Bayesian analysis. 
We present a practical implementation of a population-informed single-event analysis which relies on a double-reweighting procedure and makes use of existing data products (Sec.~\ref{sec:methods}). 
Finally, we apply our formalism to the current LIGO/Virgo catalog (Sec.~\ref{sec:results}).
We obtain population-informed posterior distributions for all the events in the catalog and consider, for the first time in GW astronomy, the correlations between event and population parameters. 
We discuss the prospects of our work in Sec.~\ref{sec:discussion} and present some generalizations in Appendices \ref{app:includingN}, \ref{app:full} and \ref{hellingersec}.  Additional results are provided as supplemental material.

Although we restrict ourselves to GW astronomy, our statistical methods are very general and can be applied to any scenario where individual observations need to be analyzed as part of a larger set while consistently accounting for their intrinsic detectability.

\section{Hierarchical Bayesian Inference} 
\label{sec:inference}
\subsection{Notation} \label{sec:notation}
Let $i\in\bigN$ label observed events in a catalog and $d_i$ denote the strain for event $i$. 
Events are described by parameters $\theta_i$ (e.g.\ BH masses and spins).
The collection of all $\Nobs$ parameters and strain data are denoted $\alltheta$ and $\alldata$ respectively.
It will be necessary to use sets with one specific event, say $j$, is omitted; these are denoted $\mosttheta$ and $\mostdata$.

We assume that we have an astrophysical population model (``$\POP$'') which depends on parameters $\lambdabar$ and predicts an expected number of sources $N(\lambdabar)$ distributed such that the number with parameters in volume $\dint\theta$ is given by
\begin{equation}
    \frac{\dint N}{\dint\theta} = N(\lambdabar) p_\POP(\theta|\lambdabar),
\end{equation}
with $\int\dint\theta\,p_\POP(\theta|\lambdabar)=1$.  
It is convenient to reparameterize $\lambdabar = (N,\lambda)$,  separating out one parameter $N$ that describes the rate from the remaining $\lambda$ that describe the shape of the population (e.g.\ the slope of the mass function and locations of any mass gaps).
With this change of variables, one has $N(\lambdabar)=N$ and $p_\POP(\theta|\lambdabar)=p_\POP(\theta|\lambda)$.

\subsection{Full hierarchical model} 
\label{sec:hierarchical}
Given all the observed data $\alldata$, what we would like to do is to simultaneously infer the properties of all the individual events $\alltheta$ and the population parameters $\lambdabar$. 
The posterior on these parameters is given by the following hierarchical Bayesian model,
\begin{align} \label{eq:hierarchical}
    P_\POP(\alltheta, \lambdabar |\alldata) = \frac{\mathcal{L}(\alldata|\alltheta)p(\alltheta|\lambdabar)\pi_\POP(\lambdabar)}{\mathcal{Z}_\POP}.
\end{align}
where $\mathcal{Z}_\POP=P_\POP(\alldata)$ is the evidence.\footnote{Hereafter, a subscript is used to denote a conditional probability. E.g.\ $P_\POP(\alldata)\equiv P(\alldata|\POP)$ is the probability of observing the data given a particular astrophysical model for the population.}

If events are independent and non-overlapping, then the first term in the numerator of Eq.~(\ref{eq:hierarchical}) is the product of the individual event likelihoods,
\begin{align} \label{eq:hierarchical_likelihood}
    \mathcal{L}(\alldata|\alltheta) = \prod_{i=1}^{\Nobs} \mathcal{L}(d_i|\theta_i) ,
\end{align}
where $\mathcal{L}(d_i|\theta_i)$ is the usual single-event GW likelihood (e.g.~\cite{1987thyg.book..330T}).

The second term describe an in-homogeneous Poisson process and involves a factor of the population model for each event (cf.\ \cite{2019MNRAS.486.1086M}) 
\begin{align} \label{eq:hierarchical_pop}
    p(\alltheta|\lambdabar) \propto e^{-N\alpha(\lambda)} N^{\Nobs} \prod_{i=1}^{\Nobs} p_\POP(\theta_i|\lambda) ,
\end{align}
where we again split $\lambdabar=(N,\lambda)$ and discard normalization terms that do not depend on $\alltheta$ or $\lambdabar$. 
The \emph{efficiency}
\begin{align} \label{eq:efficiency}
    \alpha(\lambda) = \int\mathrm{d}\theta\;P_\theta(\DET)p_\POP(\theta|\lambda) .
\end{align}
is the fraction of events in the population that are detectable and accounts for selection effects via the inclusion of a detection probability for a given event.\footnote{The detection probability $0\leq P_\theta(\DET)\leq 1$ is the conditional probability of detecting an event given its parameters; i.e.\ $P(\DET | \theta)$, hence the subscript $\theta$. Other authors write this as $p_\DET(\theta)$.}

The final term in Eq.~(\ref{eq:hierarchical}) is the Bayesian prior on the population parameters, $\pi_\POP(\lambdabar)$.

As the focus of this study is mainly on the shape of the population, not on the event rate, here we will restrict to the case where the (improper) 
prior on $N$ is scale invariant, i.e. $\pi_\POP(\lambdabar)\propto \pi_\POP(\lambda)/N$, and marginalize over $N$. A generalized derivation that includes $N$ and is suitable for inferring on the rate of events is presented in Appendix \ref{app:includingN}.
Marginalizing over $N$, the posterior in Eq.~(\ref{eq:hierarchical}) becomes
\begin{align} \label{eq:hierarchical_marg}
    P_\POP(\alltheta, \lambda |\alldata) \propto \mathcal{L}(\alldata|\alltheta)p(\alltheta|\lambda)\pi_\POP(\lambda) .
\end{align}
where
\begin{align} \label{eq:hierarchical_term_marg}
    p(\alltheta|\lambda) \propto \alpha(\lambda)^{-\Nobs} \prod_{i=1}^{\Nobs} p_\POP(\theta_i|\lambda).
\end{align}

The logical structure of this full hierarchical model is indicated graphically in Fig.~\ref{pgms}(d). This model is impractical to sample from directly due to its high dimensionality and the computational cost of its evaluation.
In practice, two of its marginalized distributions are used, as described in the following subsections.

\subsection{Population inference} \label{subsec:population_inference}
First, consider marginalizing Eq.~(\ref{eq:hierarchical}) over all the $\alltheta$.
This is what is typically done GW population inference \cite{2019ApJ...882L..24A,2021ApJ...913L...7A}
and leaves a posterior on just the population parameters
\begin{align} \label{eq:population_posterior_marg}
    P_\POP(\lambda |\alldata) &\propto \int\dint\alltheta\; P_\POP(\alltheta,\lambda|\alldata) , \\
    & \propto \frac{\pi_\POP(\lambda)}{\alpha(\lambda)^{\Nobs}}
    \prod_{i=1}^{\Nobs}\int\mathrm{d}\theta_i\;\mathcal{L}(d_i|\theta_i)p_{\rm pop}(\theta_i|\lambda) . \nonumber 
\end{align}
In order to sample this distribution, it is necessary to evaluate the $\theta_i$-integrals; in practice this is done using parameter estimation samples, see Sec.~\ref{subsec:pop}.
The logical structure of this analysis is indicated graphically in Fig.~\ref{pgms}(b).

\subsection{Population-informed single-event inference} \label{subsec:pop_informed_priors}
Second, consider marginalizing Eq.~(\ref{eq:hierarchical}) over $\lambda$ and $\mosttheta$, resulting in a posterior on just the parameters for event $j$. 
Using Eqs.~(\ref{eq:hierarchical_likelihood}), (\ref{eq:hierarchical_marg}) and (\ref{eq:hierarchical_term_marg})
and bringing the terms that depend on $\theta_j$ in front of the product on $i$ yields
\begin{align} \label{eq:pop}
    P_\POP(\theta_j&|\alldata) = \int \dint\lambda \int  \dint\mosttheta \;P_\POP(\alltheta,\lambda|\alldata)  \\
    &= \mathcal{L}(d_j|\theta_j) \int\dint\lambda\; \frac{p_\POP(\theta_j|\lambda)}{\alpha(\lambda)} P_\POP(\lambda|\mostdata) , \nonumber
\end{align}
where $P_\POP(\lambda|\mostdata)$ is given by Eq.~(\ref{eq:population_posterior_marg}) but with event $j$ omitted from the catalog; i.e.
\begin{align} \label{eq:population_posterior_marg_loo}
    P_\POP(\lambda |\mostdata) &\propto \frac{\pi_\POP(\lambda)}{\alpha(\lambda)^{\Nobs-1}}
    \prod_{i\neq j}
    \int\mathrm{d}\theta_i\;\mathcal{L}(d_i|\theta_i)p_{\rm pop}(\theta_i|\lambda) .
\end{align}
This is the typical end-product of the so-called ``{leave-one-out}'' analyses, where individual events are excluded from the catalog. 
These analyses can be used as part of a posterior-predictive test for the presence of outlying events from the main population.
For instance, some of the analyses in Refs.~\cite{2019ApJ...882L..24A,2021ApJ...913L...7A} were performed excluding either GW170729, GW190521, or GW190814.

Equation~(\ref{eq:pop}) can be rewritten in a suggestive manner:
\begin{align} \label{eq:pop_informed_prior}
    P_\POP(\theta_j|\alldata) \propto \mathcal{L}(d_j|\theta_j)\varpi(\theta_j| \mostdata) ,
\end{align}
where 
\begin{align} \label{eq:leave_one_out_prior}
    \varpi(\theta_j| \mostdata) = \int\dint\lambda\; \frac{p_\POP(\theta_j|\lambda)}{\alpha(\lambda)} P_\POP(\lambda|\mostdata) .
\end{align}
This now resembles Bayes' theorem where $\varpi$ plays the role of a \emph{population-informed prior} (which also incorporates selection effects) for the parameters of event $j$.
Crucially, this expression relies on the leave-one-out posterior $P_\POP(\lambda|\mostdata)$, thus avoiding double-counting the event $j$ in Eq.~(\ref{eq:pop_informed_prior}).
Given the properties of the $\Nobs-1$ events we have observed so far, and what we know about the sensitivity of my instrument, the {population-informed} prior $\varpi$ quantifies what we expect the $\Nobs^{\rm th}$ event to look like. 
Although we informally refer to $\varpi$ as a prior distribution, this analogy must be used carefully because the quantity $\varpi(\theta)$ is not normalized.

The logical structure of this analysis is indicated graphically in Fig.~\ref{pgms}(c). First $\Nobs-1$ events are used to learn about the population, this information is encoded in the distribution $P_\POP(\lambda|\mostdata)$. 
This is used, along with information about selection effects, to build the (pseudo) prior distribution $\varpi(\theta_j|\mostdata)$ for the parameters of the final event.
Finally, this prior is combined with the likelihood for the final event to obtain the posterior distribution $P_\POP(\theta_j|\alldata)$ on those event parameters.

It is also important to remember that population-informed reanalysis of event $j$ is necessarily conditioned on a particular model for the population.

\subsection{Empirical Bayes method} \label{subsec:empirical}
A important comment is due about the distribution $P_\POP(\theta_j|\alldata)$ in Eq.~(\ref{eq:pop_informed_prior}): care must be taken not to include event $j$ twice in the inference; it is for this reason that the integrals in Eqs.~(\ref{eq:population_posterior_marg_loo}) and (\ref{eq:leave_one_out_prior}) must be taken over the leave-one-out distribution $P_\POP(\lambda|\mostdata)$ which only use information from the \emph{other} events in the catalog. 

One could attempt to use population-informed priors \emph{without} omitting the event from the prior. 
This is a known approach to approximate the outcome of a hierarchical analysis and is often referred to as ``empirical Bayes'' \cite{casella1985introduction}.
In our context, the empirical Bayes approximation of the single-event parameters reduces to using 
\begin{align} \label{eq:popwrong}
    P_\POP&(\theta_j|\alldata) 
    \approx \mathcal{L}(d_j|\theta_j) \int\dint\lambda\;  p_\POP(\theta_j|\lambda)  P_\POP(\lambda|\alldata)
\end{align}
instead of Eq.~(\ref{eq:pop}). 
The PGM for an analysis based on this expression looks identical to that of Fig.~\ref{pgms}(c) with the exception that the event $j$ is not excluded from the box when iterating over $i$. 
We stress that Eq.~(\ref{eq:popwrong}) does not follow from the hierarchical Bayesian model in Eq.~(\ref{eq:hierarchical}) and, compared to Eq.~(\ref{eq:pop}), double counts event number $j$. 

Assuming event $j$ is not a population outlier, we would expect the empirical Bayes approximation to become increasingly accurate as the size of the GW catalog increases. 
As we will see below in Sec.~\ref{sec:results}, it seems that we are already in this regime and the empirical Bayes method generally provides a good approximation to the full hierarchical analysis. 
However, it is an important conceptual point that event $j$ is being double-counted in this analysis (it enters once in the population-informed prior and once in the likelihood) and that this is only an approximation to the full hierarchical Bayesian model. 
This conceptual point is in some ways analogous to Bessel's $N/(N-1)$ correction factor in the frequentist estimation of the variance in a population.

There have been some previous attempts at using population-informed priors in the GW context. 
The literature on this topic is rather opaque and it is often unclear which expressions are actually being used.
Careful reading of Refs.~\cite{2020ApJ...891L..31F, 2021ApJ...913L...7A, 2020ApJ...900..177K} suggests that the empirical Bayes is implicitly being used and that one event is erroneously being double-counted, although private discussions with some of the authors indicate that this is in fact not the case.
Ref.~\cite{2020ApJ...895..128M} considers a population of GW events modeled using a restricted set of parameters but does explicitly use the correct leave-one-out expression in Eq.~(\ref{eq:pop}).
The heuristic expressions for the population-informed prior reported in Refs.~\cite{2020PhRvD.102h3026G,2020ApJ...904L..26F} do not double-count any events, however they differ in the treatment of selection effects by not including the extra factor of $\alpha(\lambda)$ in Eq.~(\ref{eq:pop}); therefore, those treatments do not follow from a hierarchical Bayesian analysis, although in practice similar numerical results are obtained. 
Finally there are some unpublished technical documents \cite{DCC_Callister,DCC_Fishbach,GITHUB_Farr} which also describe the leave-one-out population-informed posterior in Eq.~(\ref{eq:pop}) together with a condition-reweighting approach to sampling this distribution which differs from the double-reweighting strategy put forward in this paper.

\subsection{Correlations between event parameters and population parameters}\label{eventpopcorr}
The full hierarchical model of Eq.~(\ref{eq:hierarchical}) contains information on the correlations between the event parameters $\alltheta$ and the population parameters $\lambda$.
These correlations are lost if one pursues the common population approach described in Sec.~\ref{subsec:population_inference} where the event parameters are marginalized over. 
The correlations are also lost when one pursues the population-informed single-event analysis described in Sec.~\ref{subsec:pop_informed_priors} (or the empirical Bayes approximation described in Sec.~\ref{subsec:empirical}) as this analysis marginalizes over the population parameters.

In order to study correlations between individual event parameters $\theta_j$ and the population parameters $\lambda$ we can marginalize the full hierarchical model of Eq.~(\ref{eq:hierarchical}) over $\mosttheta$. This gives
\begin{align} \label{eq:popcorr}
    P_\POP(\theta_j,\lambda|\alldata)
   = \mathcal{L}(d_j |\theta_j) \frac{p_\POP(\theta_j|\lambda)}{\alpha(\lambda)} P_\POP(\lambda|\mostdata) \,.
\end{align}
This distribution can be used to assess the extent to which certain parameters from specific events might be affecting the population parameters, see Sec.~\ref{subsec:results_popcorr}.

\section{Sampling and reweighting} \label{sec:methods}
A straightforward implementation of the population-informed individual-event analysis in Sec.~\ref{subsec:pop_informed_priors} would require $\Nobs$ dedicated population inference runs to be performed, one leaving out each event in turn.
This is clearly impractical, especially as $\Nobs$ becomes large. 
In this section we present a 
double-reweighting strategy that avoids the needs for these expensive calculations.

First, we briefly describe the two analyses that are commonly performed in GW astronomy: single-event parameter estimation (Sec.~\ref{subsec:PE}) and population inference (Sec.~\ref{subsec:pop}). 
We then show how the results of these can be used to obtain weighted samples from the full population-informed prior analysis in Eq.~(\ref{eq:pop_informed_prior}) via a double-reweighting strategy (Sec.~\ref{subsec:hierarchical}).

\subsection{Parameter estimation} \label{subsec:PE}
Parameter estimation (PE) is routinely performed on individual GW events using an uninformative\footnote{All priors introduce some information in the analysis. In this context ``uninformative'' indicate a prior choice that does not make use of data from the other events in the catalog.} prior, $\pi_\PE(\theta)$; for instance, common choices include priors that are uniform in masses and isotropic in spins.
The target posterior for this analysis is given by,
\begin{align} \label{eq:event_posterior}
    P_\PE(\theta_i| d_i) \propto \mathcal{L}(d_i|\theta_i)\pi_\PE(\theta_i)\,.
    \end{align}
A subscript PE indicates that a probability is conditioned on the assumptions in the uninformative prior used in single-event PE. 
The simple logical structure of this analysis indicated graphically in Fig.~\ref{pgms}(a).

An output of the PE analysis is a set of (equally weighted;\footnote{We use the notation $(x^a, w^a)\sim P(x)$, $a=1,2,\ldots,N$, to indicate weighted samples from distribution $P(x)$. The importance sampling estimate for the expectation of a function of $x$ is given by $\mathrm{E}[f(x)]=\sum_{a=1}^{N}(w^a f(x^a))/\sum_{a=1}^{N}(w^a)$.} $w^k_i=1$) posterior samples,
\begin{align} \label{eq:pe_samples}
    \big(\theta_{i}^{k}, w_i^k= 1\big)  \sim P_\PE(\theta_i|d_i),\quad\textrm{for}\;k=1,2,\ldots,S_i.
\end{align}

\subsection{Population inference} \label{subsec:pop}
Higher-level population analyses are performed with the target posterior distribution $P_\POP(\lambda|\alldata)$ described in Sec.~\ref{subsec:population_inference}.
Again, the subscript $\POP$ is there to remind us that this is a probability conditioned on the assumptions in a particular population model.
 
In practice, the samples from the PE analyses of individual events are used to efficiently evaluate the population likelihood. 
The integrals in Eq.~(\ref{eq:population_posterior_marg}) are usually approximated by the following Monte Carlo sums using Eq.~(\ref{eq:event_posterior}) and the samples in Eq.~(\ref{eq:pe_samples}),
\begin{align} 
    \int\mathrm{d}\theta_i\;\mathcal{L}(d_i|\theta)p_{\rm pop}(\theta_i|\lambda) &\propto \int\mathrm{d}\theta_i\; P_\PE(\theta_i|d_i)\frac{p_\POP(\theta_i|\lambda)}{\pi_\PE(\theta_i)} \nonumber\\
    &\approx \frac{1}{S_i}\sum_{k=1}^{S_i}\frac{p_{\rm pop}(\theta_i^k|\lambda)}{\pi_{\rm PE}(\theta_i^k)}.
    \label{eq:monte_carlo}
\end{align}
This is a reweighting procedure, where the ratio in the integrand divides by the original PE prior and multiplies by the desired population model. 

An output of this analysis (e.g.\ \cite{2021ApJ...913L...7A}), is a set posterior samples (again, equally weighted, $\omega^l=1$) drawn from the target distribution;
\begin{align} \label{eq:pop_samples}
    \big(\lambda^l, \omega^l=1\big) \sim P_{\rm pop}(\lambda|\alldata),\quad\textrm{for}\;l=1,2,\ldots,\mathcal{S}.
\end{align}

\subsection{Reweighting samples for population-informed single-event inference} \label{subsec:hierarchical}
We take the samples $\theta_j^k$ from Eq.~(\ref{eq:pe_samples}) and reweight them to the population-informed target distribution in Eq.~(\ref{eq:pop_informed_prior}) for a single event;
\begin{align} 
    \big(\theta_j^k, W^k\big) \sim P_{\rm pop}(\theta_j|\alldata).
\end{align}
The necessary weights are given by the ratio of the target distribution to the current PE distribution,
\begin{align}
    W^k = \frac{P_\POP(\theta_j^k|\alldata)}{P_\PE(\theta_j^k|d_j)}.
\end{align}
Using Eqs.~(\ref{eq:population_posterior_marg_loo}), (\ref{eq:pop_informed_prior}) and (\ref{eq:leave_one_out_prior}) for the numerator, and Eq.~(\ref{eq:event_posterior}) for the denominator, this expression simplifies to 
\begin{align}
    W^k = \frac{1}{\pi_\PE(\theta_j^k)}\int\dint\lambda\,\frac{P_\POP(\lambda|\alldata)p_\POP(\theta_j^k|\lambda)}{\int\dint\theta_j\,\mathcal{L}(d_j|\theta_j)p_\POP(\theta_j|\lambda)}.
\end{align}
The integral in the denominator is the same as that in Eq.~(\ref{eq:population_posterior_marg}) and can be approximated by the Monte Carlo sum in Eq.~(\ref{eq:monte_carlo}).
The integral in the numerator can also be be approximated by a Monte Carlo sum using the population samples in Eq.~(\ref{eq:pop_samples}). Doing so gives 
\begin{align}
    W^k = \frac{1}{\pi_\PE(\theta_j^k)\mathcal{S}}\sum_{l=1}^{\mathcal{S}}\frac{ p_\POP(\theta_j^k|\lambda^l)S_j}{\sum_{k'=1}^{S_j}\frac{p_\POP(\theta_j^{k'}|\lambda^l)}{\pi_\PE(\theta_j^{k'})}}.
\end{align}

The samples $\lambda^l$ were obtained from a likelihood that involves a reweighting of the PE samples, $\theta_i^k$.
We are now using the $\lambda^l$ samples to reweight the original PE samples $\theta_i^k$ to target the population-informed single-event posterior.
In this sense this procedure is a double reweighting of existing posterior samples.

Via an entirely analogous calculation, an additional set of weights can be obtained to approximate  $P_\POP(\theta_j,\lambda|\alldata)$ of Eq.~(\ref{eq:popcorr}) and thus capture correlations between parameters and hyperparameters.

Finally, one can also attempt a similar reweighting targeting the full hierarchical model of Sec.~\ref{sec:hierarchical} (this involves reweighting samples on $\lambda$ and $\alltheta$ simultaneously). We have attempted this and found that it is not possible to do accurately with given existing LIGO/Virgo data products. The much higher dimensionality of the distribution means that results are dominated by errors due to the finite number of samples available. Although we are reporting a null result, we provide the expressions for the full Hierarchical reweighting in Appendix~\ref{app:full} and plan to re-investigate this point in the future with different techniques.

\begin{figure*}[t]
    \includegraphics[width=\textwidth]{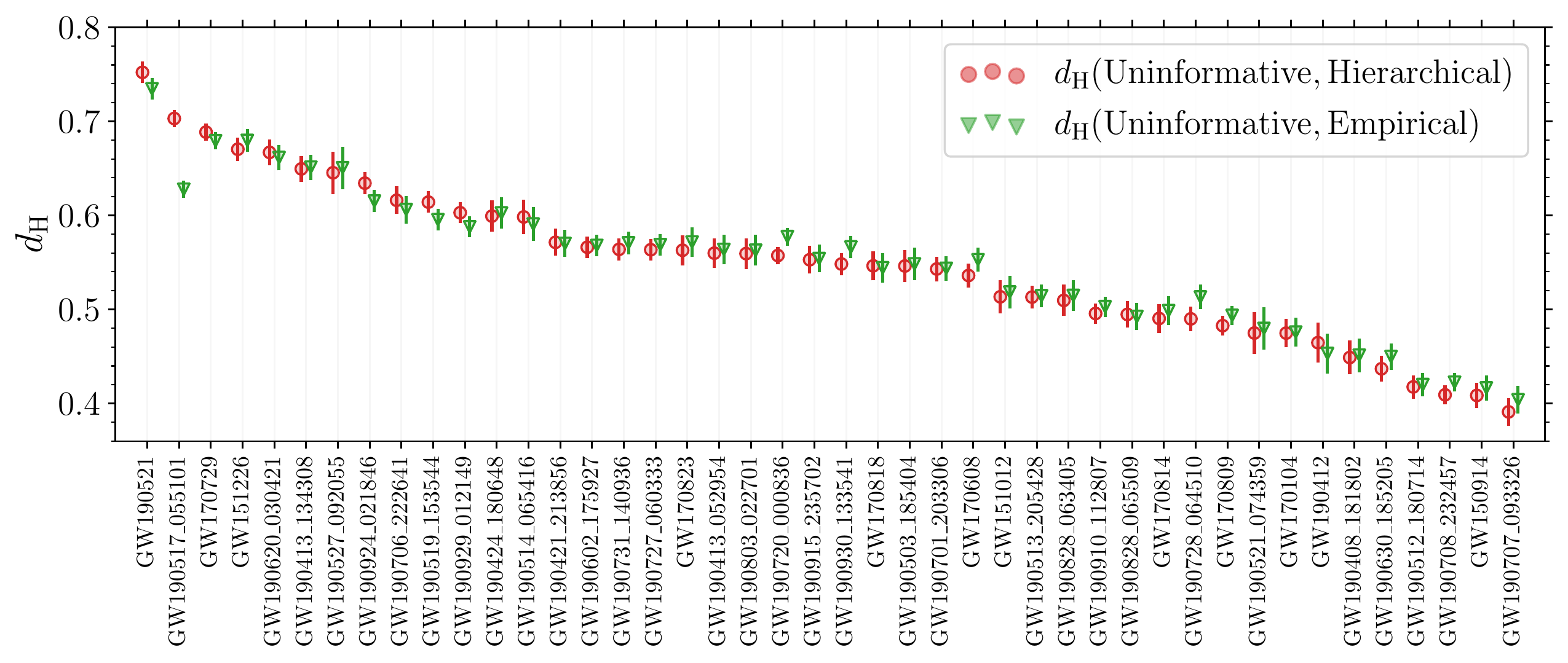}
    \caption{ \label{fig:dH}
        The Hellinger distances between the standard parameter-estimation 6-dimensional posterior distributions on $\theta_i$ (uninformative prior) and the population-informed posteriors. Results are shown for all 44 BH binary events detected with false-alarm rate $<1 {\rm yr}^{-1}$ during the first two and a half LIGO/Virgo observing runs. ``Hierarchical" (red circles) refers to the population-informed expression derived in this paper, cf Eq.~(\ref{eq:pop}); ``Empirical'' (green triangles) refers to the approximation of Eq.~(\ref{eq:popwrong}) which double-counts the event considered. Errors are estimated by splitting posterior samples into 10 sets and computing the standard deviation of $d_{\rm H}$ computed from each set.}
\end{figure*}

\section{Application to the LIGO/Virgo catalog} \label{sec:results}
We refer to the different analyses as follows:
\begin{itemize}
\item {``Uninformative''}: the posterior $P_\PE(\theta_j| d_j)$ in Eq.~(\ref{eq:event_posterior}) obtained from single-event parameter-estimation that does not use the rest of the catalog.
\item {``Hierarchical''}: the population-informed posterior $P_\POP(\theta_j|\alldata)$ of Eq.~(\ref{eq:pop}), which was obtained by marginalizing the full hierarchical model. In practice, we sample this using the double-reweighting procedure described in Sec.~\ref{subsec:hierarchical}.
\item {``Empirical''}: the empirical Bayes approximation to the population-informed posterior $P_\POP(\theta_j|\alldata)$ given in Eq.~(\ref{eq:popwrong}), where the analyzed event enter twice. This is also sampled using a double-reweighting procedure.
\end{itemize}

We use both the parameter estimation samples for $\theta$ and population inference samples for $\lambda$ provided with Ref.~\cite{2021ApJ...913L...7A}. 
This analysis includes the $\Nobs = 44$ confident binary BH events detected with false-alarm rate $<1 {\rm yr}^{-1}$ during the first two and a half observing runs of LIGO/Virgo. 
We consider individual events to be described by the following 6 parameters: $\theta=\{m_1,m_2,\chi_1,\chi_2,\theta_1,\theta_2\}$ where $m_i$ are the BH masses, $\chi_i$ are the dimensionless BH spin magnitudes, and $\theta_i$ are the tilt angles between the BH spins and the orbital angular momentum (subscripts $i=1,2$ refer to the heavier and the lighter BH, respectively). 

For our population model, we use the 
\textsc{Power Law + Peak}
model from Ref.~\cite{2021ApJ...913L...7A} as an example; however, we stress that our method is general and can be applied to any population model.
This 
\textsc{Power Law + Peak}
model depends on $\dim(\lambda)=12$ population parameters which are briefly described here.
The primary mass is assumed to follow a power-law distribution between $m_{\rm max}$ and $m_{\rm min}$ with a spectral index $\alpha$ and a Gaussian peak component designed to model a possible pileup of events below the pair-instability supernova mass gap (the peak has fractional strength $\lambda_{m}$ and is located at $\mu_{m}$ with a width $\sigma_{m}$).
The mass ratio $q=m_2/m_1\leq 1$ is assumed to follow a power-law distribution with index $\beta_{q}$.
Both the primary mass and mass ratio distributions are smoothed over a range of masses $\delta_{m}$ at the low mass end.
The dimensionless spin magnitudes are assumed to follow a beta distributions parameterised by a mean $\mu_{\chi}$ and a variance $\sigma^{2}_{\chi}$.
The distributions of the spin tilt cosines have a uniform component and a (truncated) Gaussian component with fractional strength $\zeta$ and width $\sigma_t$ that is designed to model field formation channels that might preferentially form binaries with aligned spins. 
The redshift $z$ is assumed to be distributed uniformly in comoving volume and source frame time  independently of $\lambda$.

\subsection{Catalog summary}
We compute the three posterior distributions highlighted above (``uninformative'', ``hierarchical'', and ``empirical'') for each of the 44 events. 
The complete set of posteriors for all 44 events is provided as a supplementary file.
Inspecting these plots reveals that, in general, the most noticeable effect of using a population-informed prior is to reduce the error on the mass ratio $q$ measurement and shift the posterior towards $q=1$; this is expected as most of the events in the catalog are consistent with $q=1$ and this trend is captured by the population model. 
It is also noticeable that in most cases there is excellent agreement between the ``Hierarchical'' posterior and the ``Empirical'' approximation.
Here we attempt to summarize the results for all these events by computing the Hellinger distance $d^2_{\rm H}(p,q)\equiv 1- \int\!\sqrt{p(x)q(x)} \dint x\in[0,1]$ \cite{hellinger1909neue} between the ``Uniformative'' and the other two population-informed distributions. 
A short summary of the key properties of the Hellinger distance is presented in Appendix~\ref{hellingersec}. 
Although we find the Hellinger distance to be a useful indicator, we caution against overinterpreting of these results; it is not a substitute for visually inspecting the distributions.

Figure \ref{fig:dH} shows the distances between the uninformative distribution and the (i) hierarchical and (ii) empirical distributions. 
The resulting values of $d_{\rm H}$ are in the range $\sim 0.4$--$0.7$, indicating that the population-informed reweighting has a relative major impact on the interpretation of the systems.
The events with largest $d_{\rm H}$ have either large masses (GW190521, GW170729, GW190929\_012149) or large effective spin (GW151226, GW190517\_055101, GW190620\_030421) compared to the rest of the catalog~\cite{2019PhRvX...9c1040A,2021PhRvX..11b1053A}. 
In general, the events with low $d_{\rm H}$ (i.e. those for which the population-informed reweighting is less impactful) have either parameters that are well in the bulk of the population or large signal-to-noise ratios such that the posterior is mostly driven by the likelihood and not the prior. GW150914 is an excellent example of this.

We find that the two distances we computed (uninformative vs hierarchical and uninformative vs empirical) are, in general, very similar to each other with differences as small as  $\Delta d_{\rm H}\lesssim 0.01$. 
This is within the errors on $d_{\rm H}$, which we estimate by splitting the samples weights into 10 subsets and computing the resulting standard deviations. Overall we find that, despite being an important conceptual point, the Bessel-like correction highlighted in this paper is subdominant. Some exceptions include GW190517\_055101 ($\Delta d_{\rm H}\sim 0.07$, see Sec.~\ref{spinofspinnyguy} below),  GW190728\_064510, GW190720\_000836, and GW190924\_021846 ($\Delta d_{\rm H}\sim 0.02$)

Let us now focus on two cases of particular interest. 

\subsection{The masses of GW190521}

\begin{figure}
    \includegraphics[width=\columnwidth]{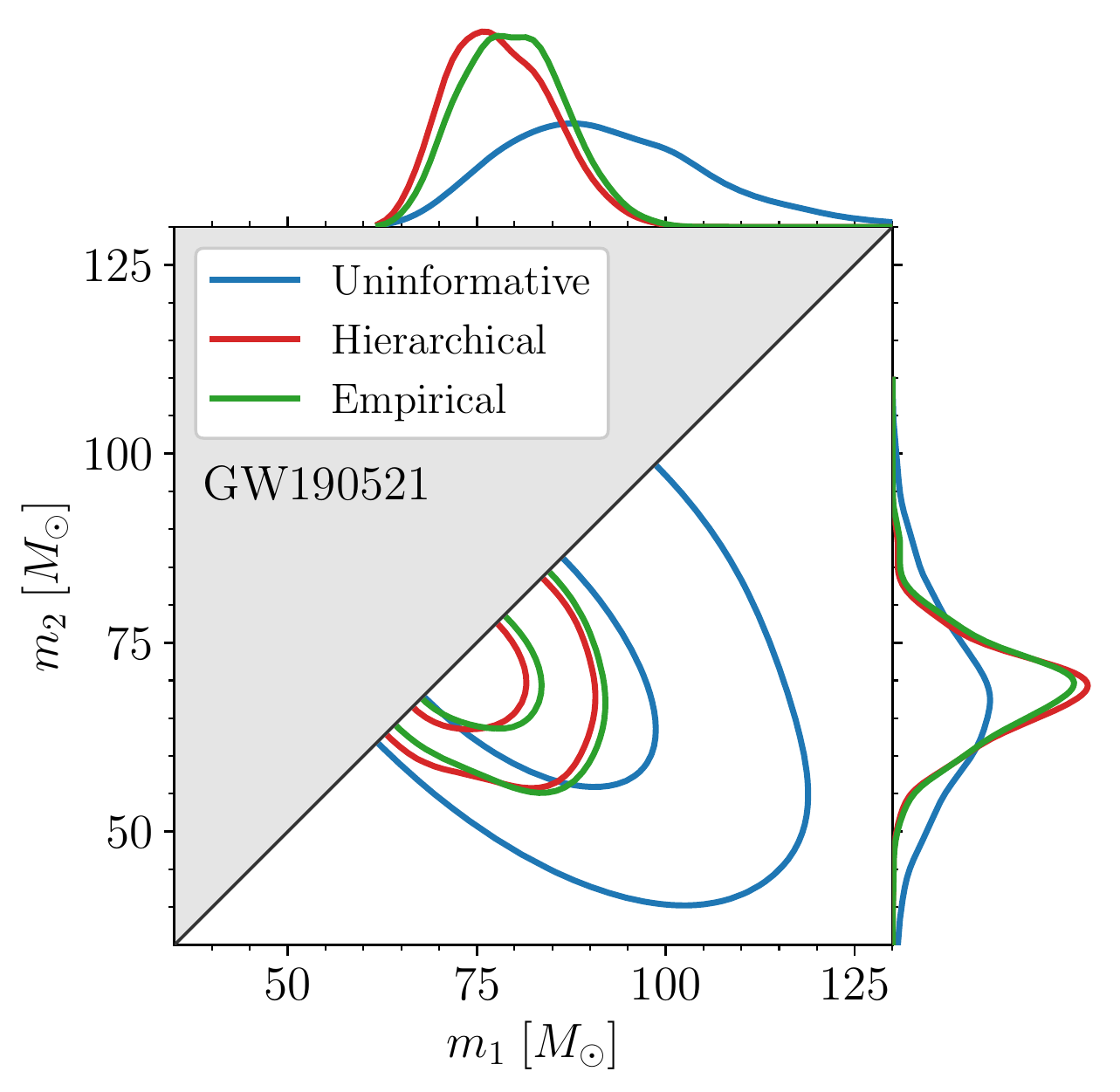}
    \caption{ \label{fig:GW190521_masses}
        The posteriors for the source-frame component masses of the heaviest binary observed to date, GW190521. 
        The blue contours show results of the single-event parameter estimation, [``Uninformative'', Eq.~(\ref{eq:event_posterior}].
        The red contours show the population-informed single-event inference developed in this paper [''Hierarchical'', Eq.~(\ref{eq:pop})].
        The green contours show the empirical Bayes approximation to the population-informed single-event distribution [''Empirical'', Eq.~(\ref{eq:popwrong})].
        We show 50\% and 90\% contours.
    }
\end{figure}

GW190521 is the heaviest binary in the catalog.
It is also the event for which the use of a population-informed prior makes the biggest difference to the posterior in the sense that it has the largest value of $d_{\rm H}$ in Fig.~\ref{fig:dH}.
Figure~\ref{fig:GW190521_masses} shows the posterior distributions on its component masses $m_1$ and $m_2$. 
Priors that are inferred from the (rest of the) population shift the posterior distribution toward lower masses compared to the uninformative prior adopted in the single-event parameter-estimation analysis (cf. Ref.~\cite{2021ApJ...913L...7A} for an analogous discussion). 
This is unsurprising because all of the other events have masses that are likely to be lower than those of GW190521. 
We find that the hierarchical and empirical estimate of the population-informed posterior return essentially the same result: it appears that, already with a catalog of $\sim 50$ events, the double counting of one event involved in the empirical Bayes approximation has a small effect on the result. 
This is compatible with the results the posterior-predictive checks reported in Ref.~\cite{2021ApJ...913L...7A}, which concluded that GW190521 is not a population outlier. 

\begin{figure}
    \includegraphics[width=\columnwidth]{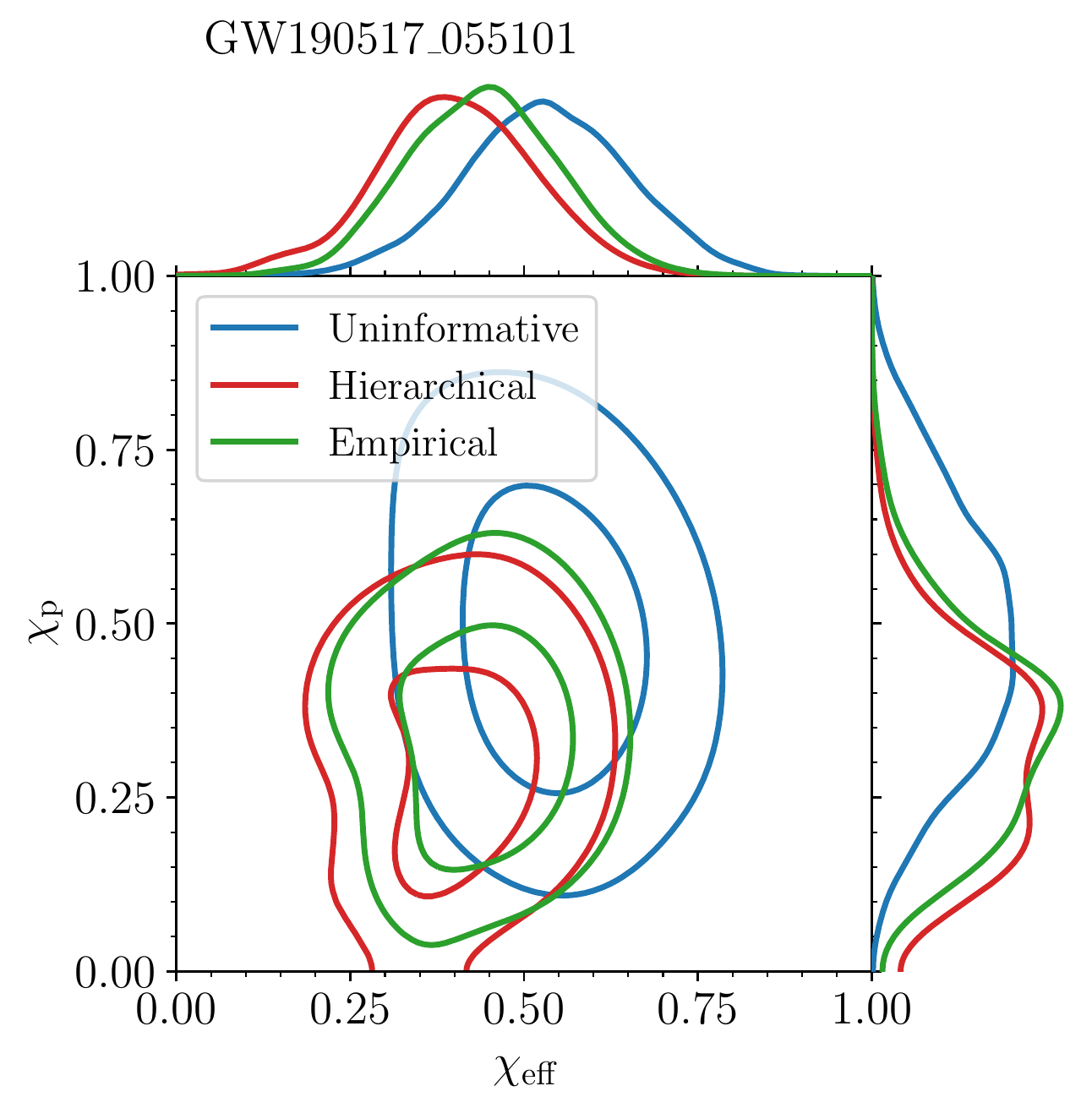}
    \caption{ \label{fig:GW190517_055101_spins}
        The posteriors for the spin combinations $\chi_{\rm eff}$ and $\chi_{\rm p}$ of the binary with the largest $\chi_{\rm eff}$ in the catalog, GW190517\_055101. 
        The blue, red and green contours show the results of the ''Uninformative'' [Eq.~(\ref{eq:event_posterior}], ''Hierarchical'' [Eq.~(\ref{eq:pop})], and ''Empirical'' [Eq.~(\ref{eq:popwrong})] inferences respectively.         We show 50\% and 90\% contours.
        }
\end{figure}

\subsection{The spins of GW190517\_055101} \label{spinofspinnyguy}

GW190517\_055101 is the event with the largest effective spin $\chi_{\rm eff}$  \cite{2008PhRvD..78d4021R} in the catalog (this is the spin combination parameter that is currently best measured) and also present a moderate value of $\chi_{\rm p}$ \cite{2015PhRvD..91b4043S} (another measured spin combination that captures the precession of the orbital plane; see Ref.~\cite{2021PhRvD.103f4067G} for issues and refinements on its definition). 
Figure~\ref{fig:GW190517_055101_spins} illustrates the joint distribution of these two parameters for GW190517\_055101. Much like for the masses of GW190521, GW190517\_055101 has higher effective spins compared to the rest of the catalog and thus the population-informed posteriors peak at lower values compared to the uninformative ones. In this case, the difference between the full hierarchical estimate and the empirical approximation is more pronounced. 
When the event is double counted (empirical Bayes method), the high-spin value of GW190517\_055101 contaminates the population-informed prior and produces a distribution with $\chi_{\rm eff}=0.44^{+0.17}_{-0.18}$ (median and 90\% credible interval). 
The result obtained from marginalizing the full hierarchical model instead returns $\chi_{\rm eff}=0.40^{+0.17}_{-0.19}$, which is further from the uninformative results $\chi_{\rm eff}=0.53^{+0.19}_{-0.20}$. 
However, although the maximum-a-posteriori is affected by the erroneous double counting, the difference between the peaks of the hierarchical and empirical of $\chi_{\rm eff}$ posteriors is still smaller than the widths of the distributions. 
This suggests that, at least at the present signal-to-noise-ratio, the empirical Bayes approximation is appropriate. 
The population-informed estimates of $\chi_{\rm p}$ peak at lower values compared to the uninformative case; this is driven by the bulk of the observed catalog which shows relatively little evidence for spin precession \cite{2021ApJ...913L...7A}.

\subsection{Correlations between event and population parameters} \label{subsec:results_popcorr}

Beside re-investigating single events in light of the population, our double reweighting approach allows us to study the correlations between the parameters of single events (i.e. the $\theta$'s) and those of the population (i.e. the $\lambda$'s). 
This is the first time that such $\theta$--$\lambda$ correlations are studied in the published GW literature (although we note some 
unpublished results are present at~Ref.~\cite{DATA_RELEASE}).
Studying these correlations requires a suitable marginalization of the full hierarchical model as described in Sec.~\ref{eventpopcorr} and cannot be done with the usual approach of, e.g., Refs.~\cite{2019ApJ...882L..24A,2021ApJ...913L...7A}.

\begin{figure}
      \includegraphics[width=\columnwidth]{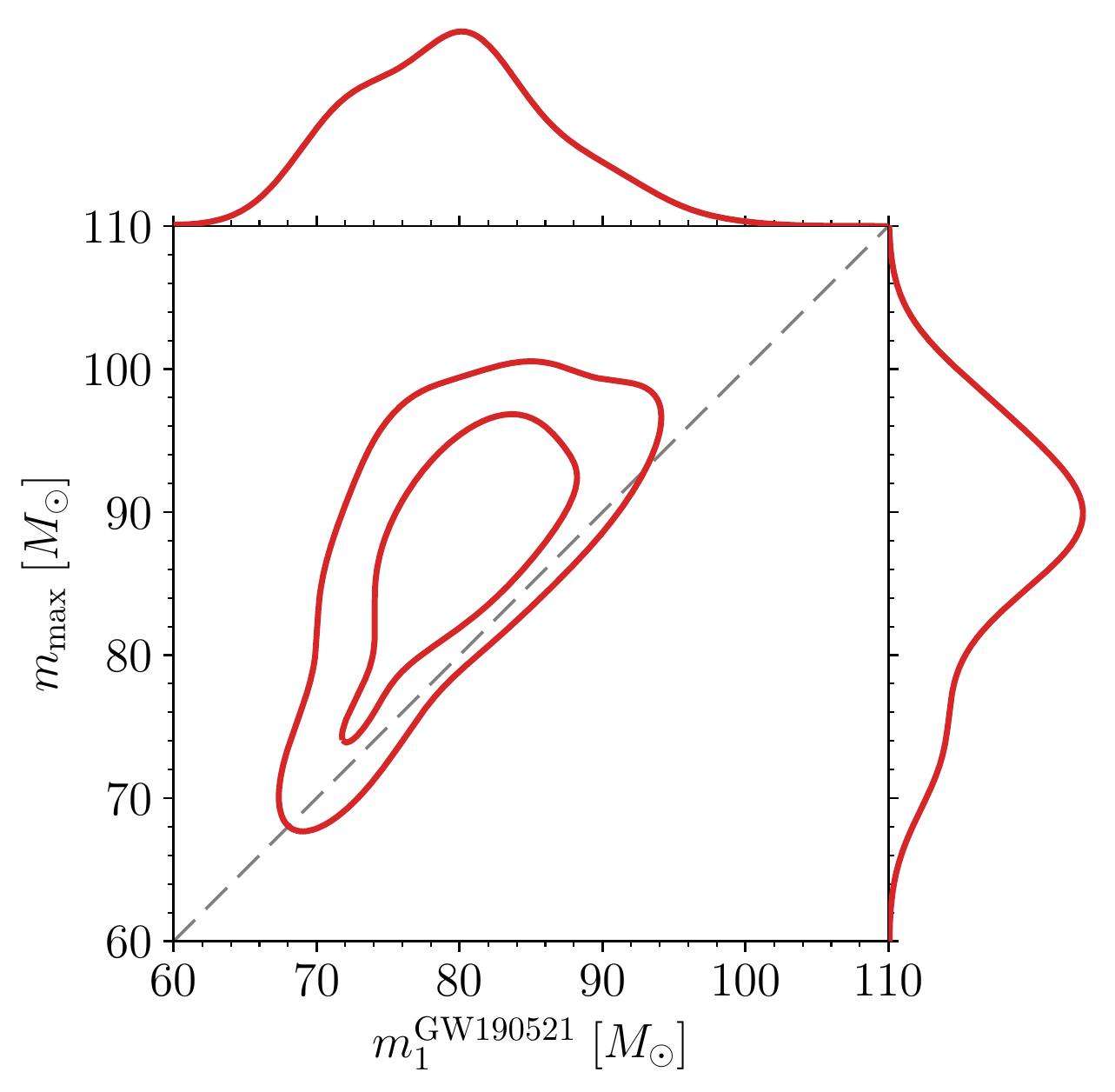}
    \caption{Correlations between the primary mass of GW190521 (one of the event parameters $\theta$) and the maximum mass cutoff in the source distribution (one of the population parameters $\lambda$). Investigating $\theta$--$\lambda$ correlations requires considering the full hierarchical model (cf. Sec.~\ref{eventpopcorr}) and cannot be tackled with the most common population approach which marginalizes over the event parameters (cf. Sec.~\ref{subsec:population_inference} and \ref{subsec:pop}).}
    \label{thetalambda}
\end{figure}

Figure~\ref{thetalambda} shows the joint distribution of the primary mass $m_1$ of GW190521 (which is part of $\alltheta$) and the upper mass cutoff in the population model $m_{\rm max}$ (which is instead part of $\lambda$). 
Among the many combinations between parameters and hyperparameters, this choice is particularly interesting because the value of the upper mass cutoff in the population fit is largely driven by the need to accommodate the largest mass in the catalog (although, the 
\textsc{Power Law + Peak}
population model we are using can accommodate a fraction of events above $m_{\rm max}$, depending on the values of the ``peak'' parameters $\lambda_m$, $\mu_m$, and $\sigma_m$). 
Indeed, we find that the two quantities are moderately strongly correlated. This result can be read both ways: (i) if the mass of GW190521 turns out to be on the lower edge of its credible interval, one infers a lower mass cutoff in the entire population; (ii) if statistical and systematic errors in the population inference cause an overestimate of the mass cutoff (i.e. $m_{\rm max}$ is smaller than what comes out of the inference), then the population-informed estimate of GW190521 points to lighter BHs.

This is just one of the possible analyses that can be performed with the method we presented. Going forward, we argue these correlations should be investigated in greater detail as they might shed further light on the GW data-analysis procedure and the related astrophysical interpretation of the sources.

\section{Discussion}\label{sec:discussion}
Inference on single events and inference on the  population are part of a common Bayesian hierarchical model. These two aspects are, in practice, tackled individually for practical and computational constraints. In the LIGO context, this is made possible by the fact that sources are not overlapping, such that one can take a short time segment of data containing an event and analyze it independently on the rest of the  data stream. This will not be possible with next-generation GW detectors. Source overlap will require the development of the so-called ``global fit'', tackling individual events, populations, and noise all at the same time (this is especially true for LISA, but 3rd generation ground-based detectors will also be affected).

But even for current detector networks, developing a full hierarchical model might uncover new features in the data. This paper tackles a particular aspect of this problem, namely the characterization of individual events in light of the (rest of the) detected population, an approach which is often referred to as using a ``population-informed prior''. The key result is presented in Eqs.~(\ref{eq:pop_informed_prior}) and (\ref{eq:leave_one_out_prior}).

We presented a thorough derivation of population-informed single-event statistics in the presence of selection effects, which we hope clarifies implicit assumptions and inconsistencies present in previous treatments. In particular, we highlighted the conceptual difference between the correct marginalization stemming from the full hierarchical model and the so-called empirical Bayes method, where the targeted event is effectively double counted.

Our formalism has several applications.
Firstly, one can constrain the properties of GW events under the plausible assumption that they belong to a common population of sources. While we find that the conceptual correction pointed out here is subdominant, we stress it can be applied ``for free'', using data products that are routinely produced and made available. We thus argue that this correction should be applied to all GW analyses that make use of population-informed priors.
Secondly, this approach further allows us to study, for the first time in the context of GW astronomy, the correlations between the parameters of individual events and those of the population. The line of investigation that is put forward in this paper has the potential to unveil new details on the (astro)physics of GW sources from existing and future data.

Selection effects ---analogous to the so-called Malmquist bias in observational astronomy--- play an important role in GW population inference \cite{2019PASA...36...10T,2020arXiv200705579V}. For population-informed single-event analyses like those presented here, selection effects turn out to be a nuisance. In this context, one is interested solely in the \emph{observed} population of sources, not the \emph{observable} one. Our formalism contains selection effects because we wish to rely on current pipelines (e.g. Refs.~\cite{2019ApJ...882L..24A,2021ApJ...913L...7A}) where $\lambda$ parameterizes the intrinsic population of BHs. This is also indicated in the PGM of Fig.~\ref{pgms}(c), where selection effects are first introduced when inferring $\lambda$ from $\mosttheta$ and then removed when reconstructing $\theta_j$ from $\lambda$. We stress that developing a simpler population analysis without selection effects ---targeting the observed population and not the intrinsic one---, can also have applications in GW astronomy, an example of which is presented here. If the observed population has parameters $\tilde\lambda$, our equations can be immediately applied by substituting $\lambda\to\tilde \lambda$ and setting $\alpha(\lambda)=1$.

We hope that the formalism presented in this paper can set the stage for deeper explorations of GW data exploiting the interplay between single events and populations.

\begin{acknowledgments}
    We thank 
    Riccardo Buscicchio, 
    Tom Callister, 
    Guy Davies, 
    Reed Essick, 
    Will Farr, 
    Maya Fishbach, 
    Alexander Lyttle, 
    Matthew Mould, 
    Colm Talbot, 
    and
    Alberto Vecchio
    for discussions.
    D.G. is supported by European Union's H2020 ERC Starting Grant No. 945155---GWmining, Leverhulme Trust Grant No. RPG-2019-350, and Royal Society Grant No. RGS-R2-202004. Computational work was performed on the University of Birmingham BlueBEAR cluster.
\end{acknowledgments}


\appendix

\section{Rates analysis} \label{app:includingN}

In this appendix, we present the more general expressions one obtains without marginalizing over the expected number of events $N$. 

The starting point is given by Eq.~(\ref{eq:hierarchical_pop}). 
Marginalizing over the individual-event parameters as in Sec.~\ref{subsec:population_inference} yields
\begin{align} 
    P_\POP(\lambdabar|\alldata) &\propto \pi_\POP(\lambdabar) e^{-N \alpha(\lambda)} N^{\Nobs}
     \nonumber \\ &\times \prod_{i=1}^{\Nobs}\int\mathrm{d}\theta_i\;\mathcal{L}(d_i|\theta_i)p_{\rm pop}(\theta_i|\lambda).
\end{align}
With a calculation analogous to that of Sec.~\ref{subsec:pop_informed_priors},  the population-informed single-event posterior can be obtained by  instead marginalizing Eq.~(\ref{eq:hierarchical_pop}) over $\theta_{i\neq j}$ and $\lambdabar$. We obtain
\begin{align} 
    P_\POP(\theta_j|\alldata) &\propto \mathcal{L}(d_j|\theta_j)\varpi(\theta_j| \mostdata ,
\end{align}
where 
\begin{align} 
    \varpi(\theta_j| \mostdata) =\int\dint\lambdabar\;  N\,p_\POP(\theta_j|\lambda)  P_\POP(\lambdabar|\mostdata)
\end{align}
and
 \begin{align} 
 P_\POP(\lambdabar|\mostdata) & \propto  \pi_\POP(\lambdabar) e^{-N \alpha(\lambda)} N^{\Nobs-1} \\
    &\times
    \prod^{\Nobs-1}_{i\neq j}  \int\dint\theta_i   \;\mathcal{L}(d_i|\theta_i) p_\POP(\theta_i|\lambda)\,. \notag
\end{align}

\section{Full hierarchical reweighting} \label{app:full}
Here we highlight a possible strategy to combine samples from single-event and population analysis to target the full hierarchical posterior of Eq.~(\ref{eq:hierarchical}). We have attempted to implement this strategy and report here that this approach fails due to the limited numbers ($\sim 10^3$-$10^4$) of samples provided in public LIGO/Virgo data products.

Let us take one sample from each of the individual PE posterior chains (Eq.~[\ref{eq:pe_samples}]) and a sample from the population inference posterior chain (Eq.~[\ref{eq:pop_samples}]) and denote the result $X^m=(\alltheta^m, \lambda^m)$.
This combined sample follows a distribution  $Q(\alltheta,\lambda)$ that is the product of $\Nobs+1$ target distributions of the individual analyses, i.e.
\begin{align}
    Q(\alltheta,\lambda) &\propto P_\POP(\lambda|\alldata) \prod_{i=1}^{\Nobs}P_\PE(\theta_i| d_i) .
\end{align}

Posterior samples from the full hierarchical model in Eq.~(\ref{eq:hierarchical}) can be obtained by a reweighting of $X^m$;
\begin{align}
    \big(X^m, W^m\big) \sim P_\POP(\alltheta,\lambda|\alldata) ,\quad\textrm{for}\;m=1,2,\ldots,s,
\end{align}
where $s=\mathrm{min}(\mathcal{S}, \{S_i\})$. 
The necessary weights are given by the ratio of the target distribution in Eq.~(\ref{eq:hierarchical}) to the $Q$-distribution evaluated at the $X^m$ samples: 
\begin{align}
    W^m &= \frac{P_\POP(\alltheta^m,\lambda^m|\alldata)}{Q(\alltheta^m,\lambda^m)} ,  \\
    &= \prod_{i=1}^{\Nobs} \frac{p_\POP(\theta_i^m|\lambda^m)}{\pi_\PE(\theta_i^m)}\left[\int\dint\theta_i\;\mathcal{L}(d_i|\theta_i)p_\POP(\theta_i|\lambda^m)\right]^{-1} \notag \\
    &\approx \prod_{i=1}^{\Nobs} \frac{p_\POP(\theta_i^m|\lambda^m)S_i}{\pi_\PE(\theta_i^m)}\left[\sum_{k=1}^{S_i}\frac{p_\POP(\theta_i^{k}|\lambda^m)}{\pi_\PE(\theta_i^{k})}\right]^{-1} . \notag
\end{align}
\vspace{0.1cm}

\vspace{-0.5cm}

\section{Properties of the Hellinger distance} \label{hellingersec}

There are many ways to quantify the difference between probability distributions (see, for example, \cite{Measures_of_Information}).
There is no particularly natural choice for our problem.
Nevertheless, we find it convenient to quote a single number to quantify the \emph{distance} between two distributions; we use the Hellinger distance \cite{hellinger1909neue} for this purpose.

The Hellinger distance between distributions $p(x)$ and $q(x)$ is defined as
\begin{align}
    d_{\rm H}^2(p,q) = 1-\int\dint x\,\sqrt{p(x)q(x)}
\end{align}
and has a few convenient properties.
First, it is symmetric: $d_{\rm H}(p,q) = d_{\rm H}(q,p)$. It also lies  the range $0 \leq d_{\rm H}\leq 1$ with the equalities occurring when $p$ and $q$ are identical or disjoint (mutually exclusive), respectively.
To gain intuition for what a value of $d_{\rm H}$ ``means'', one can compute the distance between offset Gaussians: if $p=\mathcal{N}(\mu,\Sigma)$ and $q=\mathcal{N}(\mu + c,\Sigma)$ then $d_{\rm H}^2(p,q) = 1-\exp(-c\cdot\Sigma^{-1}\cdot c/8)$ in any dimension.

Finally, we present a toy Bayesian calculation that is useful for gaining further intuition about $d_{\rm H}$.
Suppose we infer the value of parameters $x\in\mathbb{A}$.
We have a prior $\pi(x)$, normalized such that $\int_\mathbb{A}\dint x\,\pi(x) = 1$.
We make observations with a likelihood function $\mathcal{L}(\mathrm{obs}|x)\propto \mathbbm{1}_\mathbb{B}(x)$, where $\mathbbm{1}$ is the indicator function and $\mathbb{B}\subset \mathbb{A}$.  That is, our observations exclude with certainty some values of $x$ while allowing all others with equal likelihood. 
Using Bayes' theorem, the posterior distribution is $P(x|\mathrm{obs})=\mathbbm{1}_\mathbb{B}(x)\pi(x)/ F$, where the normalization $F=\int_\mathbb{B}\dint x\,\pi(x)$ is the fraction of the prior allowed by our observations. 
We find that the squared Hellinger distance between our prior and posterior is related to this fraction by $d^2_{\rm H}(P,\pi)=1-\sqrt{F}$. 

\nocite{suppl}

\bibliographystyle{apsrev4-2} 
\bibliography{refs}

\end{document}